\documentclass[aps,%
prd,%
twocolumn,%
nofootinbib,%
noshowkeys,%
superscriptaddress,%
longbibliography,%
floatfix,%
preprintnumbers]{revtex4-1}
\usepackage[utf8]{inputenc} 
\usepackage{hyperref} 
\usepackage{amsmath} 
\usepackage{amsfonts}
\usepackage{amsthm}
\usepackage{amssymb}
\usepackage{graphicx} 
\usepackage{slashed} 
\usepackage{color} 
\usepackage{ulem}

\graphicspath{
	{./}
  {./figs/}
}

\newcommand{\mr}[1]{\mathrm{#1}}
\newcommand{\mb}[1]{\mathbf{#1}}
\newcommand{\mc}[1]{\mathcal{#1}}

\newcommand{\F}{\sqrt{\mb{B}^2-\mb{E}^2}}
\newcommand{\plong}{\mb{p}_\parallel}
\newcommand{\ptrans}{\mb{p}_\perp}

\definecolor{ARcolour}{rgb}{0,0.5,0}

\newcommand{\Helsinki}{\affiliation{
    Department of Physics and Helsinki Institute of Physics,
    PL 64, 
    FI-00014 University of Helsinki,
    Finland
}}
\newcommand{\Nottingham}{\affiliation{
    School of Physics and Astronomy,
    University of Nottingham,
    Nottingham NG7 2RD,
    United Kingdom
}}
\newcommand{\Imperial}{\affiliation{
    Department of Physics,
    Imperial College London,
    London SW7 2AZ,
    United Kingdom
}}

\begin{document}

\title{Schwinger pair production of magnetic monopoles:\\
momentum distribution for heavy-ion collisions}
\date{March 26, 2021}

\author{Oliver Gould}
\email{oliver.gould@nottingham.ac.uk}
\Nottingham
\Helsinki

\author{David L.-J. Ho}
\email{d.ho17@imperial.ac.uk}
\Imperial

\author{Arttu Rajantie}
\email{a.rajantie@imperial.ac.uk}
\Imperial

\preprint{HIP-2021-12/TH, IMPERIAL-TP-2021-DH-05}

\begin{abstract}
Magnetic monopoles may be produced by the dual Schwinger effect in strong magnetic fields.
Today, the strongest known magnetic fields in the universe are produced fleetingly in heavy-ion collisions.
We use the complex worldline instanton method to calculate the momentum distribution of magnetic monopoles produced in heavy-ion collisions,
in an approximation that includes the effect of the magnetic field to all orders but neglects monopole self-interactions.
The result saturates the preparation time-energy uncertainty principle, and yields a necessary ingredient for experimental monopole searches in heavy-ion collisions.
\end{abstract}

\maketitle

\section{Introduction} \label{sec:introduction}

The search for magnetic monopoles has a long history, going back at least to Petrus Peregrinus who, in the 13th century, searched for isolated magnetic poles in fragments of lodestone.
In modern particle physics, the search is motivated by
the possibility to explain the quantisation of electric charge~\cite{Dirac:1931kp,Polchinski:2003bq},
and because magnetic monopoles are predicted by broad classes of models, including Grand Unified Theories~\cite{tHooft:1974kcl,Polyakov:1974ek,Preskill:1984gd} and semiclassical gravity with topology change~\cite{Gibbons:1990um,Maldacena:2020skw}.
More recently, there has been a revival of interest in models containing low mass solitonic magnetic monopoles, which may in principle be probed by terrestrial collider experiments~\cite{cho2013finite, ellis2016price, arunasalam2017electroweak, hung2020topologically}.
Further, monopoles may exist as elementary particles, in which case their mass would be a free parameter.

Interpreting the results of collider searches for magnetic monopoles require theoretical predictions of cross sections;
experiment and theory should go hand-in-hand.
However, theory is lagging sorely behind, chiefly because of the strong coupling of magnetic monopoles~\cite{Dirac:1931kp}.
The magnetic charge must be an integer multiple of the Dirac charge $g_D=2\pi/e$, which is inversely related to the fundamental electric charge $e$.
This implies that perturbation theory, based on the loop expansion of Feynman diagrams, completely fails for magnetic monopoles.

A loophole to this dilemma exists, in that calculations can be performed at strong coupling for certain semiclassical production processes.
In particular, monopole production via the dual Schwinger effect is computationally tractable.
This is the electromagnetic dual of the usual Schwinger effect~\cite{Schwinger:1951nm}, in which a strong magnetic field decays via quantum tunnelling to form a magnetic monopole-antimonopole pair.
The probability of pair production per unit spacetime volume is given by~\cite{Affleck:1981ag,Affleck:1981bma,Ho:2021uem}:
\begin{equation} \label{eq:affleck_manton}
P \sim \exp\left(-\frac{\pi m^2}{gB} + \frac{g^2}{4}\right), 
\end{equation}
where $B$ is the magnetic field strength, $m$ is the monopole mass and $g$ is the magnetic charge.
Here and henceforth we use units where $\hbar = c = 1$ unless otherwise stated.
Note that Eq.~\eqref{eq:affleck_manton} is only valid when the process is exponentially suppressed.
For weak magnetic fields, $B\ll 4\pi m^2/g^3$, this suppression is strong, and conversely the strongest magnetic fields give the highest probability of forming magnetic monopole-antimonopole pairs.
For magnetic fields larger than $\sim 4\pi m^2/g^3$, while Eq.~\eqref{eq:affleck_manton} is no longer quantitatively reliable, monopole production is nevertheless expected to be unsupressed. This is supported by field theory instanton calculations~\cite{Ho:2021uem}.
Today the strongest known magnetic fields are produced fleetingly in ultraperipheral heavy-ion collisions~\cite{Huang:2015oca}.

Thus far there has only been one search for magnetic monopoles in heavy-ion collisions, from 1997, carried out at SPS at a centre-of-mass energy per nucleon of $\sqrt{s_{\rm NN}}\approx 17.4~\mr{GeV}$~\cite{He:1997pj}.
From the negative result of this search, upper bounds were inferred on the production cross section of magnetic monopoles in heavy-ion collisions at this centre-of-mass energy.
In Ref.~\cite{Gould:2017zwi}, by considering monopole production by magnetic fields%
\footnote{
Also considered was the boost to the production by the thermal energy of the quark-gluon plasma.
This effect is crucial at low centre-of-mass energies, but is irrelevant at LHC energies because the temperature of the plasma grows very slowly with centre-of-mass energy~\cite{Baier:2000sb}.
}
this experimental result was used to place lower bounds on the mass of possible magnetic monopoles.
The resulting lower bounds are the strongest lower bounds on the mass of magnetic monopoles which do not suffer from the inapplicability of perturbation theory to monopoles.
The bounds are also independent of the specific underlying UV theory, as they depend only on the coupling of magnetic monopoles to photons at long distances, which is fixed by Maxwell's equations and the Dirac quantisation condition~\cite{Dirac:1931kp}.

Much higher centre-of-mass energies are routinely reached in heavy-ion colliders today: approximately 200~GeV per nucleon at RHIC~\cite{Harrison:2003sb} and 5020~GeV per nucleon at the LHC~\cite{Jowett:2019jni}.
Thus, possible future searches for magnetic monopoles in heavy-ion collisions have the potential to massively extend the mass reach beyond what was achieved at SPS.
However, for this to be possible it is necessary to have theoretical control of the monopole production cross section at these higher energies.

In this paper we consider the momentum distribution of monopoles produced by the Schwinger mechanism.
This is a crucial ingredient for experimental searches, both in terms of the angular distribution and the momentum magnitude distribution, as it is a necessary input to calculate the acceptance of a given experimental setup.
For example, for a monopole to be registered in the MoEDAL trapping detectors~\cite{MoEDAL:2016jlb,Acharya:2019vtb} it must be fast enough to punch through the beam pipe, and yet be slow enough to be trapped in the aluminium rods.

In Ref.~\cite{Gould:2019myj} we considered the computation of the total cross section of monopole production in high energy heavy-ion collisions.
The task was to overcome two theoretical hurdles:
the strong coupling of monopoles to photons,
and the strong space-time dependence of the external electromagnetic fields.
Approaching this task stepwise, we utilised two different approximations:
the locally constant field approximation (LCFA), in which the strong photon-monopole coupling is accounted for to all orders, but the space-time dependence of the external electromagnetic field is treated perturbatively,
and the free-particle approximation (FPA), in which the space-time dependence of the external electromagnetic field is accounted for to all orders, but the photon-monopole coupling is treated perturbatively.
In a limited region of parameter space, we were also able to include both the strong coupling and the space-time dependence nonperturbatively by numerically solving a two-dimensional integro-differential equation, however we were not able to reach the energy scales relevant for the LHC.

Starting from the LCFA and treating the space-time dependence of the electromagnetic field perturbatively, we found that while the spatial dependence of the field reduces the cross section polynomially, the time dependence of the field increases the cross section exponentially, thus overall there is a significant boost.
Physically, this boost to particle production is due to the non-adiabaticity of time evolution in the presence of the rapidly changing electromagnetic field.
Conversely, starting from the FPA and treating the monopole-photon coupling perturbatively, we found that monopole-photon interactions increased the cross section exponentially.
Physically, this is due to the lower threshold energy for producing a monopole-antimonopole bound state which then disassociates, than for producing free particles.
Thus, both the monopole-photon coupling and the space-time dependence of the electromagnetic field increase the monopole production cross section.
As a consequence, we expect the total cross sections computed in both the LCFA and the FPA to underestimate the true cross section.
This is important for monopole searches, as it means that these cross sections give a lower bound on the expected number of monopoles produced, and correspondingly a lower bound on the monopole mass if there is no detection.

In the following section, we set up our notation and coordinate system for studying heavy-ion collisions, and review the electromagnetic fields produced in them.
In Sec.~\ref{sec:fpa} we perform a detailed calculation of the momentum distribution in the FPA, utilising the worldline instanton method, and in Sec.~\ref{sec:limitations} we assess the validity and limitations of the results.
In Sec.~\ref{sec:uncertainty}, utilising only very general arguments, we present consequences and constraints regarding the momentum distribution of monopoles.
The results of these general arguments are in accordance with those of Sec.~\ref{sec:fpa}.
Finally, in Sec.~\ref{sec:conclusions} we summarise our results and discuss the remaining theoretical uncertainties in light of upcoming experimental searches.
The momentum distribution as calculated in the LCFA is presented in Appendix~\ref{appendix:lcfa}, and the total cross sections in both the LCFA and FPA are collected in Appendix~\ref{appendix:total_cross_sections}.

\section{Setup} \label{sec:setup}

\begin{figure}[t]
    \centering
    \includegraphics[width=0.5\textwidth]{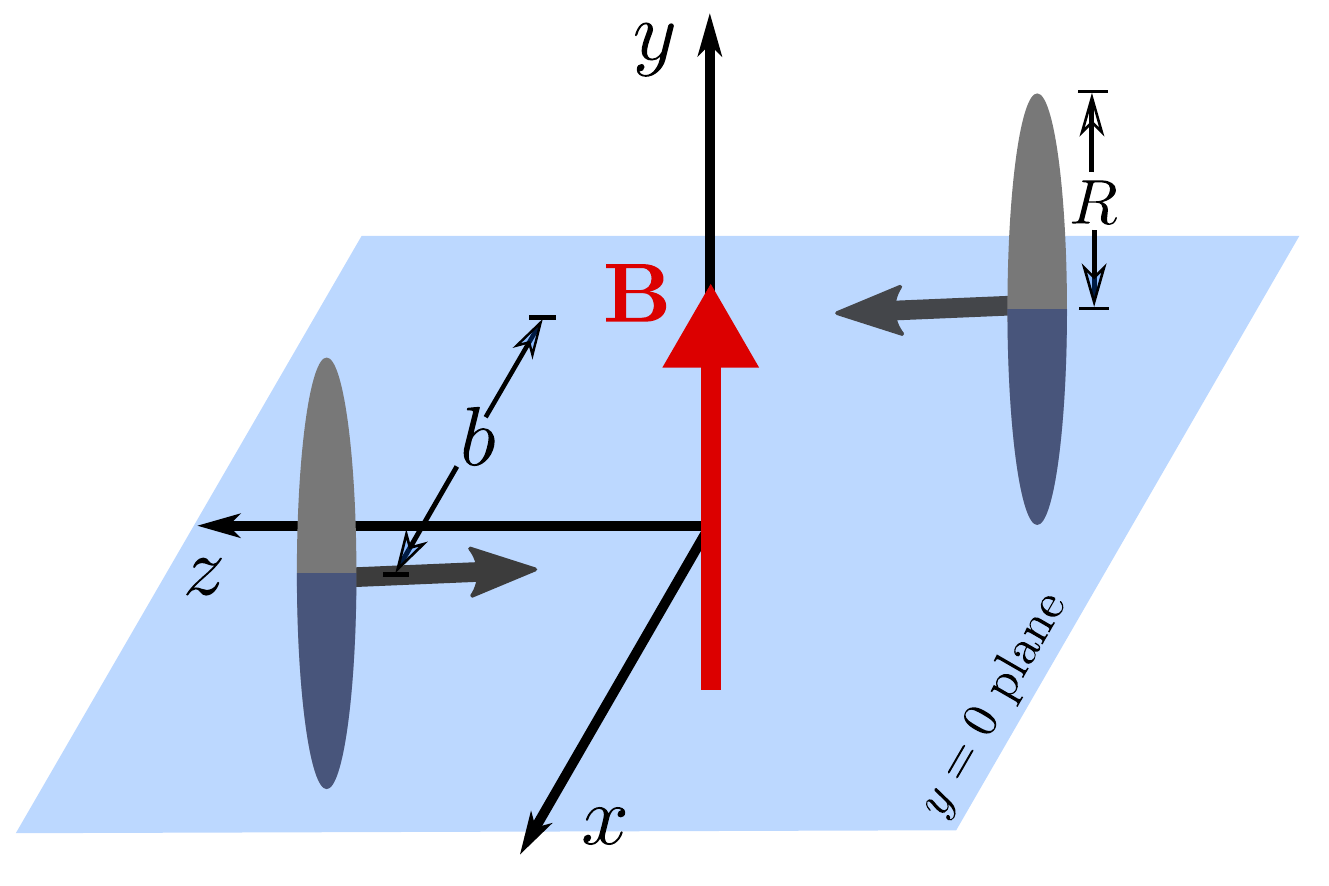}
    \caption{A heavy-ion collision described in our coordinate system, with $z$ the beam axis and the impact parameter $b$ lying along the $x$-axis.
    For ultraperipheral collisions $b\approx 2R$.
    At the centre of the collision, the electric fields of the two ions cancel, while the magnetic fields reinforce.
    }
    \label{fig:coordinates}
\end{figure}

Ultraperipheral heavy-ion collisions lead to the strongest magnetic fields.
In these collisions the two ions skim past each other with the magnitude of the impact parameter $b$ equal to approximately twice the nuclear radius $R$,
leaving a relatively empty detector signal.%
\footnote{
Interestingly ultraperipheral heavy-ion collisions have also been studied in order to constrain Born-Infeld theory~\cite{Ellis:2017edi} and consequently the magnetic monopoles predicted by that theory.
}
The magnetic fields produced are strongly spacetime-dependent.
Fig.~\ref{fig:coordinates} sets up our coordinate system:
$z$ is the beam axis, along which the ion velocities lie and
$x$ is chosen for each collision to be the direction along which the impact parameter $\mb{b}=b \hat{\mb{x}}$ lies.
The origin of the coordinate system is the spacetime point at the centre of the collision, with $t=0$ being the time of closest approach, and as a consequence $y=0$ defines the reaction plane.

In the rest frame of an ion, its electromagnetic field is the Coulomb electric field.
Boosted to ultrarelativistic speeds, the ion and its field are Lorentz contracted along the beam axis.
A strong magnetic field is produced, according to Amp\`{e}re's law, which circulates around the ion.
Superposing the field of two ions undergoing an ultraperipheral collision, the peak of the electromagnetic scalar invariant $\sqrt{B^2 - E^2}$ occurs at the origin of our coordinates, at which point the magnetic field is nonzero and points in the $y$ direction, while the electric field is zero.
This peak field varies over a region of size $R$ in the $x$ and $y$ directions, and a region of size $R/\gamma$ in the $z$ and $t$ directions.
Here $\gamma\gg 1$ is the Lorentz factor, equal to $\gamma=1/\sqrt{1-v^2}$, where $v$ is the 3-velocity.
For a review of the electromagnetic fields produced in heavy-ion collisions, see Ref.~\cite{Huang:2015oca}.

In Ref.~\cite{Gould:2019myj}, we computed the electromagnetic fields produced by heavy-ion collisions, assuming a classical Saxon-Woods charge distribution.
In the vicinity of the origin, the numerically-computed fields were found to be well described by the following fit,
\begin{align}
B_y &= \frac{B/2}{\left[1+\omega^2(t-z)^2\right]^{3/2}}+\frac{B/2}{\left[1+\omega^2(t+z)^2\right]^{3/2}}, \nonumber \\
E_x &= \frac{B/2}{\left[1+\omega^2(t-z)^2\right]^{3/2}}-\frac{B/2}{\left[1+\omega^2(t+z)^2\right]^{3/2}}, \label{eq:fit_functions}
\end{align}
where, for ultraperipheral collisions, the two fitting parameters were given approximately by%
\footnote{
There are overall numerical factors $\approx 1$ missing from Eqs.~\eqref{eq:Bmaxgamma} and \eqref{eq:omegamaxgamma} which can be found in Ref.~\cite{Gould:2019myj} for the case of lead ions.
We have also dropped factors of $v\approx 1$, as the expressions hold only for ultrarelativistic collisions.
}
\begin{align}
 B &\approx \frac{Ze \gamma}{2\pi R^2},\label{eq:Bmaxgamma} \\
 \omega &\approx \frac{\gamma}{R}.\label{eq:omegamaxgamma}
\end{align}
For $5.02~{\rm TeV}$ lead-lead collisions at the LHC, $\gamma\approx 2675$, $B\approx 8~{\rm GeV}^2$ and $\omega\approx 80~{\rm GeV}$.
All other components of the electromagnetic field are subdominant, being suppressed relatively by powers of $1/\gamma$.
We have also dropped the much weaker $x$ and $y$-dependence of the field.
Note that while Eq.~\eqref{eq:fit_functions} is only a fit to the full result, it is nevertheless an exact solution to Maxwell's equations in vacuum.

Effects due to the spatial distribution of nucleons within the ions, as well as the finite electrical conductivity of the ions, will modify this picture in the near vicinity of the origin.
However, these cannot modify the asymptotic tails of the electromagnetic field, which are fixed by the result for point charges.
Statistical fluctuations of $\mc{O}(1)$ nucleons will cause relative perturbations of order $\mc{O}(1/Z)$, where $Z$ is the ion-charge.
In fact, due to the exponential dependence of the production cross section on the magnetic field, event-by-event fluctuations will only increase the monopole yield of a collider run.
The contribution of quantum fluctuations to the electromagnetic field is expected to be small~\cite{Deng:2012pc,Danhoni:2020ezq}, though there remains some debate.

While we expect Eq.~\eqref{eq:fit_functions} to provide a reasonable approximation to the electromagnetic fields in ultrarelativistic heavy-ion collisions, many of our conclusions in this paper will not depend on this specific functional form.
In fact, the crucial characteristics of the field for the momentum distribution of monopole production are the timescales and lengthscales on which it varies.
These are all fixed by the geometry of ultraperipheral collisions.

\section{Monopole momenta in the free particle approximation} \label{sec:fpa}
In the free particle approximation (FPA) to the production of magnetic monopoles, the effect of the external electromagnetic field (that produced by the heavy ions) is assumed to dominate over the effect of monopole-antimonopole interactions.
To leading order in this approximation, the coupling between monopoles and the external field is treated exactly, whereas interactions between the monopoles themselves are dropped.
The particle production rate and resulting momentum distribution can thus be found by solving the Dirac or Klein-Gordon equations for the mode functions in the presence of the external field.
For sufficiently slow pair production, this calculation is semiclassical, and instanton methods are applicable.

Magnetic monopoles couple to the dual electromagnetic potential \(A_\mu\), which is related to the electromagnetic field tensor by $\tfrac{1}{2}\epsilon_{\mu\nu\rho\sigma}F^{\rho\sigma}=\partial_\mu A_\nu - \partial_\nu A_\mu$, where $\epsilon_{\mu\nu\rho\sigma}$ is the Levi-Civita symbol.
We use the mostly minus Lorentz signature.
A dual electromagnetic potential which leads to the electromagnetic fields of Eq.~\eqref{eq:fit_functions} is
\begin{equation} \label{eq:Adual_full}
    A_y = 
    \frac{(B/2) \left(t-z\right)}{\sqrt{1+\omega ^2 \left(t-z\right)^2}}
    +\frac{(B/2) \left(t+z\right)}{\sqrt{1+\omega ^2 \left(t+z\right)^2}}
    \;,
\end{equation}
all other components zero.
In the context of QED, the momentum spectrum of electron-positron pairs produced by similarly spacetime dependent fields as Eq.~\eqref{eq:Adual_full} has been studied in Refs.~\cite{Hebenstreit:2011wk,Kohlfurst:2017git,Aleksandrov:2017mtq,Lv:2018wpn,Kohlfurst:2019mag,Aleksandrov:2019ddt}.
Within the FPA, the methods used in these references are directly applicable to monopole production, and their application would be valuable.

To compute the longitudinal momentum spectra of particles produced by Schwinger production, we adopt the method of Ref.~\cite{dumlu2011complex} utilising complex worldline instantons.
This involves finding periodic solutions to the equations of motion for particles in a given external field, with complex spacetime coordinates. This method is particularly useful because on the one hand it may be used to analyse temporally and/or spatially inhomogeneous external fields, and on the other hand it can be extended to incorporate an arbitrary coupling constant~\cite{Affleck:1981bma,Gould:2017fve}, moving beyond the FPA (not attempted in this work).

In Ref.~\cite{Gould:2019myj}, we argued that despite the complicated spacetime dependence of the electromagnetic fields in heavy-ion collisions, due to the symmetry under $z\to -z$, the exponential dependence of the total monopole pair production probability is the same as that for the much simpler field with only spacetime dependence through the time coordinate:
\begin{equation} \label{eq:heavyIonMagneticField}
    B_\mathrm{ext}^i = \frac{B}{[1 + (\omega t)^2]^{3/2}} \delta^{i 2},
\end{equation}
where \(B\) and \(\omega\) are the same constants as above.%
\footnote{
See also Ref.~\cite{Torgrimsson:2016ant}, where the same conclusion was reached for fields which are antisymmetric in directions transverse to the field.
}

In the following, we will use Eq.~\eqref{eq:heavyIonMagneticField} to compute the momentum distribution of Schwinger-produced monopoles in the FPA.
This spatially homogeneous field reproduces the time dependence of \eqref{eq:fit_functions}, which we have shown in Ref.~\cite{Gould:2019myj} is the most important aspect of the spacetime inhomogeneity for the total production probability. 
It is, however, spatially homogeneous, whereas \eqref{eq:fit_functions} varies strongly in the \(z\) direction.
While this does not affect the exponential dependence of the total production probability, it is expected to affect the momentum distribution.
In Section \ref{sec:limitations}, we argue that the resulting modification to the momentum distribution is at most an \(O(1)\) factor.

The relation of the dual potential to the magnetic field is \(B^i_\mathrm{ext} = \partial^i A_0 - \partial_0 A^i\).
For the magnetic field of Eq.~\eqref{eq:heavyIonMagneticField}, without $z$-dependence, the dual gauge field reduces to
\begin{equation} \label{eq:Adual_simple}
    A_\mu = \frac{B t}{\sqrt{1 + (\omega t)^2}} \delta_{\mu 2}.
\end{equation}

The worldline instanton method with imaginary \(t\) coordinate (i.e.\ a Wick rotation) was used in Ref.~\cite{Gould:2019myj} to compute the production probability of monopole-antimonopole pairs at zero initial momentum in the field given by Eq.~\eqref{eq:heavyIonMagneticField}.

In this paper, we instead follow the approach of Ref.~\cite{dumlu2011complex}: rather than Wick rotating, we promote the spacetime coordinates to complex numbers and search for periodic solutions to the equations of motion with imaginary proper time.

These equations are simply the Lorentz force law for magnetically charged particles:
\begin{equation} \label{eq:eoms}
\begin{split}
    m \ddot{t} &= \frac{g B \dot{y}}{[1 + (\omega t)^2]^{3/2}}, \\
    m \ddot{y} &= \frac{g B \dot{t}}{[1 + (\omega t)^2]^{3/2}}, \\
    m \ddot{x} &= m \ddot{z} = 0.
\end{split}
\end{equation}
Here a dot denotes a derivative with respect to proper time \(\tau\) along the particle worldline, and \(g\) is the magnetic charge of the particle. The first integral of these equations provides the constraint
\begin{equation} \label{eq:properTimeFirstIntegral}
    \dot{t}^2 - \dot{x}^2 - \dot{y}^2 - \dot{z}^2 = 1,
\end{equation}
motivating the description of \(\tau\) as proper time even when it takes an imaginary value. From the three spatial translation symmetries, Noether's theorem gives the conserved charges
\begin{equation}
\begin{split}
    p_x &= m\dot{x}, \\
    p_y &= m\dot{y} - g A_y, \\
    p_z &= m\dot{z},
\end{split}
\end{equation}
which may be interpreted as the canonical momenta of the produced particles. Eq.~\eqref{eq:properTimeFirstIntegral} can therefore be rewritten as
\begin{equation} \label{eq:firstIntegral}
    \dot{t}^2 - \frac{1}{m^2}\left[p_x^2 + \left(p_y + g A_y(t)\right)^2 + p_z^2 \right] = 1. 
\end{equation}

In order to determine the worldline trajectory, we must specify the initial values \(x_\mu(\tau = 0)\) and \(\dot{x}_\mu(\tau = 0)\). Due to the translational symmetry of the magnetic field, we are free to fix the spatial position of the worldline, and without loss of generality we can choose
\begin{equation}
x(0) = y(0) = z(0) = 0.
\end{equation}
The initial conditions on the proper time derivatives of the spatial coordinates are equivalent to specifying canonical momenta:
\begin{equation} \label{eq:initialMomenta}
    \begin{split}
        m \dot{x}(0) &= p_x, \\
        m \dot{y}(0) &= p_y + gA_y(t(0)), \\
        m \dot{z}(0) &= p_z.
    \end{split}
\end{equation}

For fields of the form \eqref{eq:heavyIonMagneticField}, the final kinetic momentum is related to the canonical momentum by
\begin{equation} \label{eq:kineticMomentum}
    k_2 = p_y + g A_y(t \to \infty) = p_y + \frac{g B}{\omega} \approx p_y + 2.4 n \ \mathrm{GeV},
\end{equation}
with the canonical and kinetic momenta being equal in other directions.
Here \(n\) denotes the number of Dirac charge quanta the monopole carries.
For the approximate numerical value of $gB/\omega$, we have assumed ultrarelativistic lead-ion collisions, as relevant to the LHC.

The remaining initial conditions to be chosen are \(t(0)\) and \(\dot{t}(0)\). These are specified by stipulating~\cite{dumlu2011complex} that the classical worldline trajectories pass through WKB turning points, defined by
\begin{equation} \label{eq:wkbTurningPoints}
    m^2 + p_x^2 + \left(p_y + gA_y(t_\mathrm{wkb})\right)^2 + p_z^2 = 0.
\end{equation}
Note that because \(t\), and thus \(A_y\), is complex, this does not imply that \(p_x = p_y = 0\). Solving this equation gives a complex conjugate pair of turning points in the complex \(t\) plane; the worldline solution interpolates between them. We thus choose the initial condition
\begin{equation}
    t(0) = t_\mathrm{wkb}.
\end{equation}
The condition \eqref{eq:firstIntegral} requires the final boundary condition
\begin{equation}
    \dot{t}(0) = 0.
\end{equation}

The fact that Eq.~\eqref{eq:wkbTurningPoints} is satisfied at \(\tau = 0\) allows a simplification of the \(\dot{y}\) initial condition: substituting Eq.~\eqref{eq:wkbTurningPoints} into Eq.~\eqref{eq:initialMomenta} gives
\begin{equation}
    m \dot{y}(0) = \pm i m_\perp,
\end{equation}
where the ``transverse mass'' is defined
\begin{equation}
    m_\perp^2 = m^2 + p_x^2 + p_z^2.
\end{equation}
The sign indicates the direction in which the worldline is traversed and does not affect the value of the action.

The probability of producing particles with a given canonical momentum \(p_y\) is given by the imaginary part of the effective action of the worldline solving Eqs.~\eqref{eq:eoms} for imaginary proper time.
\begin{equation} \label{eq:imAction}
    n(m, \mathbf{p}) \sim \exp\left(-\mathrm{Im}(S_\mathrm{eff}[t]) \right),
\end{equation}
where \cite{dumlu2011complex}
\begin{align}
    S_\mathrm{eff}[t] =& \frac{1}{2}\int_0^T \bigg\{m + m \dot{t}^2 
    \nonumber \\
    &+ \frac{1}{m}\left[p_x^2 + \left(p_y + gA_y(t)\right)^2 + p_z^2 \right] \bigg\} \mathrm{d} \tau,
\end{align}
\(T\) being the imaginary worldline period. Using Eq.~\eqref{eq:firstIntegral} and the relation \(m \dot{y}(0) = p_y + gA_y\), one can show that on shell,
\begin{equation}
    S_\mathrm{eff}[t] = m \int_0^T \dot{t}^2 \, \mathrm{d} \tau.
\end{equation}

As a final observation, note that the physical mass \(m\) factors out of the equations of motion and can be removed entirely by rescaling \(\tau \to \tau / m\). The only mass dependence in the action originates from the transverse mass term in the initial condition for \(\dot{y}\), Eq.~\eqref{eq:initialMomenta}. This means that the action at arbitrary transverse momentum can be obtained by solving the equations of motion for \(p_x = p_z = 0\), and substituting \(m \to m_\perp\). In the following, we drop the transverse momentum terms for brevity.

For \(p_y = 0\) the equations of motion \eqref{eq:eoms} are analytically solvable: the solution has \(t\) purely imaginary and takes the form of an ellipse in the \(\mathop{\mathrm{Im}}(t)\)-\(\mathop{\mathrm{Re}}(y)\) plane~\cite{Gould:2019myj}. The ellipse has semimajor and semiminor axes
\begin{align}
    a_t &= \frac{m}{gB} \frac{1}{\sqrt{1 + \xi^2}}, \\
    a_y &= \frac{m}{gB} \frac{1}{1 + \xi^2},
\end{align}
where \(\xi=m \omega / (g B)\) is the Keldysh parameter, and the imaginary part of the action is \cite{dunne2006prefactor}
\begin{equation} \label{eq:zeroMomentumAction}
    \mathop{\mathrm{Im}}(S[x_\mu]) = \frac{\pi m^2}{g B} \frac{4 [\mathbf{E}(-\xi^2) - \mathbf{K}(-\xi^2)]}{\pi \xi^2}.
\end{equation}
Here \(\mathbf{E}\) and \(\mathbf{K}\) denote elliptic integrals. As previously discussed, for ultrarelativistic heavy-ion collisions the \(\xi \gg 1\) limit is relevant: in this case
\begin{align}
    a_t &\approx \frac{m}{gB\xi} = \frac{1}{\omega}, \\
    a_y &\approx \frac{m}{gB\xi^2} = \frac{1}{\xi \omega}, \\
    \mathop{\mathrm{Im}}(S[x_\mu]) &\approx \frac{4 m^2}{g B \xi} = \frac{4 m}{\omega}.
\label{eq:zeromomapprox}
\end{align}

The elliptical worldline solution becomes increasingly prolate with increasing \(\xi\); for very large values of \(\xi\) the worldline barely deviates from the imaginary \(t\) axis.

\begin{figure}[ht]
    \includegraphics[width=0.5\textwidth]{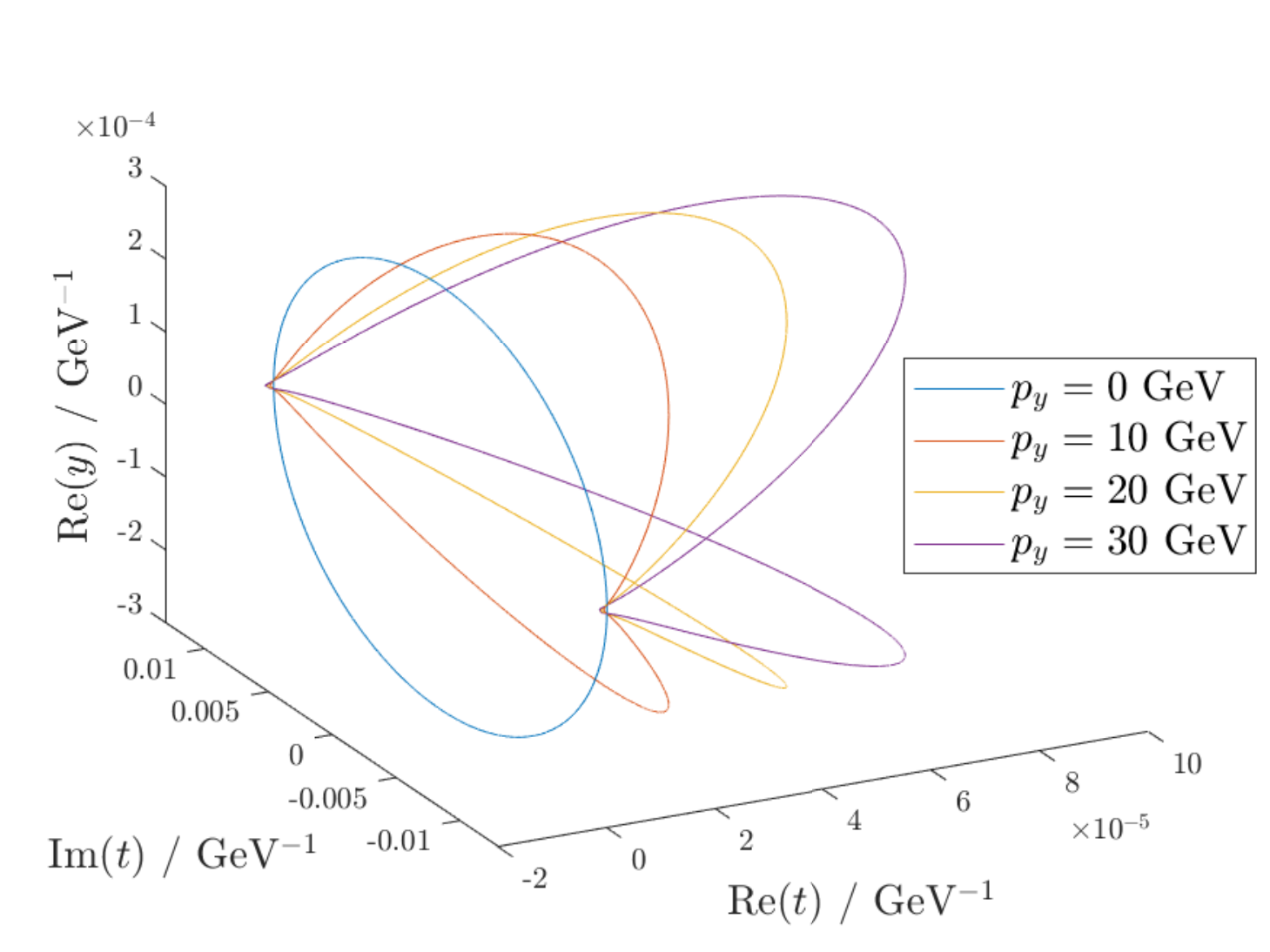}
    \caption{3D plots of complex worldline instantons in the field given by Eq.~\eqref{eq:heavyIonMagneticField}, for monopoles of mass \(100 \ \mathrm{GeV}\) and a collision energy of \(5.02 \ \mathrm{TeV}\) per nucleon.
    }
    \label{fig:complexWorldlines}
\end{figure}

For \(p_y \neq 0\), the initial condition~\eqref{eq:wkbTurningPoints} means that \(t\) is no longer purely imaginary; in Ref.~\cite{dumlu2011complex} the solutions were termed ``complex worldline instantons''. These solutions are not obtainable analytically, but can be determined using a numerical prescription outlined in Ref.~\cite{dumlu2011complex}; we have carried out this calculation for monopoles produced in collisions at LHC energies of \(5.02 \, \mathrm{TeV}\) per nucleon. The effect of nonzero longitudinal momentum is to bend the worldline away from the imaginary \(t\) axis: this is illustrated in Fig.~\ref{fig:complexWorldlines}. Note that these worldlines are not symmetric about the real \(t\) axis; worldlines with negative values of longitudinal momentum would bend in the other direction.

\begin{figure}[ht]
    \includegraphics[width=0.5\textwidth]{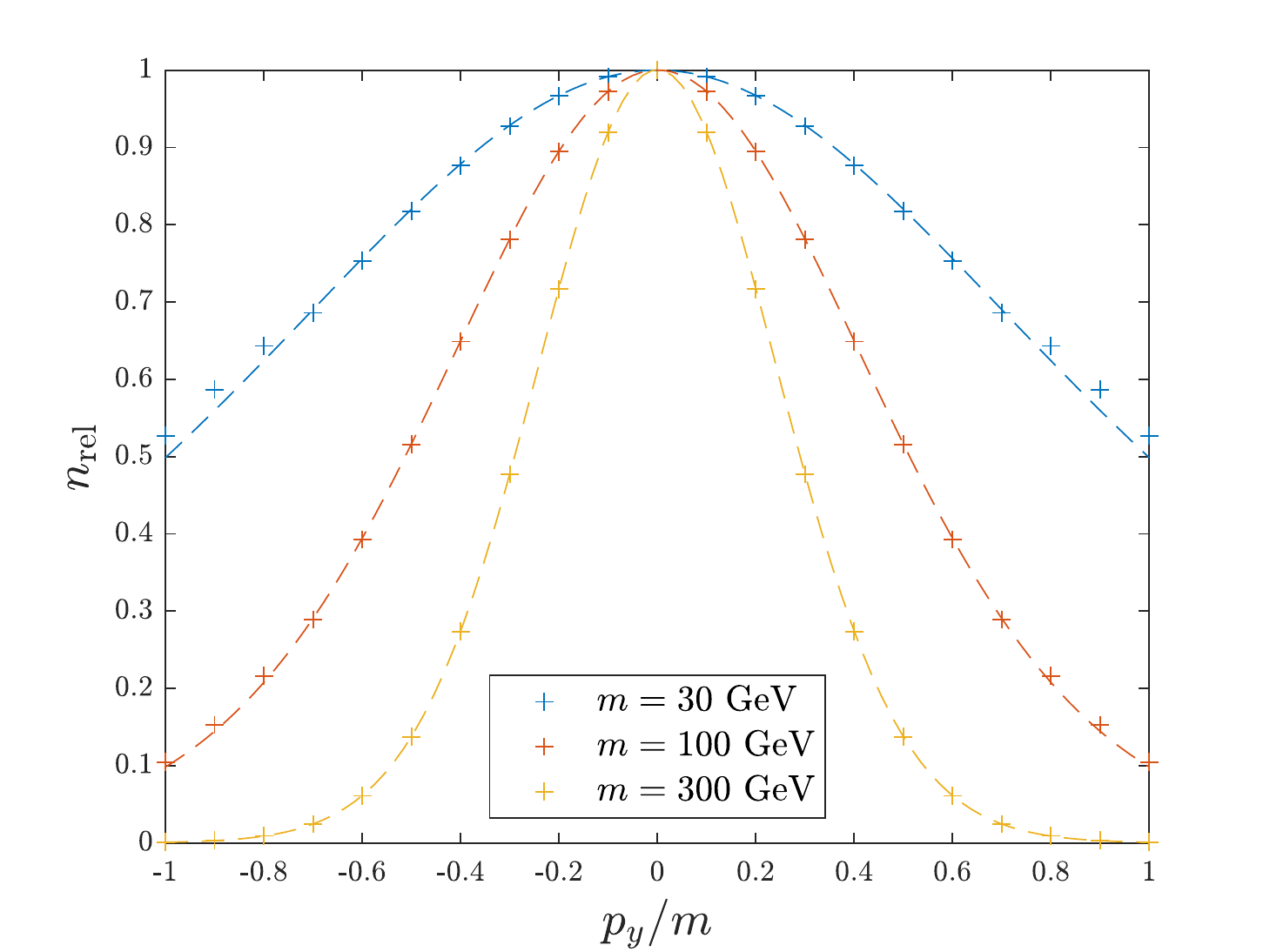}
    \caption{
    Relative momentum distribution of monopole-antimonopole pairs produced from the field \eqref{eq:heavyIonMagneticField}, for collision energies of \(5.02 \ \mathrm{TeV}\) at various monopole masses. Dashed lines show 
    Eq.~(\ref{eq:final_result}) for comparison.    
    }
    \label{fig:momentumDistributionLHC}
\end{figure}

The momentum spectrum resulting from our numerical calculation is plotted in Fig.~\ref{fig:momentumDistributionLHC}. It can be seen that the probability distribution is well approximated by the expression
\begin{equation} \label{eq:analyticMomentumDistribution}
    n(m,p_y) \sim \exp \left( -\frac{4}{\omega} \sqrt{m^2 + p_y^2} \right),
\end{equation}
which can be obtained by substituting \(m \to \sqrt{m^2 + p_y^2}\) into the zero-momentum result \eqref{eq:zeroMomentumAction} and taking the high-inhomogeneity limit.
It therefore reproduces the known analytic result at zero momentum.
Note that this is not a numerical fit, as there are no free parameters.
We conjecture that this relationship is valid for any field of the form \eqref{eq:heavyIonMagneticField}, provided that the Keldysh parameter
\begin{equation} \label{eq:highMassCondition}
    \frac{m \omega}{g B} \approx \frac{m}{2.4 n \ \mathrm{GeV}} \gg 1,
\end{equation}
where \(n\) is the number of Dirac charge quanta the monopole carries. Comparing the computed and approximated values for monopole production at LHC energies with momenta $p_y \in [-m/2, m/2]$ we find that Eq.~\eqref{eq:analyticMomentumDistribution} is accurate to within around 1\% for $m = 30 \ \mathrm{GeV}$, and is even smaller for higher masses. Excluding a narrow window not yet excluded by existing mass bounds~\cite{Gould:2017zwi}, realistic monopole models will satisfy inequality~\eqref{eq:highMassCondition}.

As the transverse momentum affects the final result only via a modification of the effective mass, the relative momentum distribution is thus the isotropic distribution
\begin{equation} \label{eq:final_result}
    n_{\rm rel}(m, \mathbf{p})\equiv
    \frac{n(m, \mathbf{p}) }{n(m, 0) }=
        \exp\left[-\frac{4}{\omega} \left(\sqrt{m^2 + |\mathbf{p}|^2}-m\right) \right].
\end{equation}

The kinetic momentum \(\mathbf{k}\) measured by a detector is related to the canonical momentum \(\mathbf{p}\) by Eq.~\eqref{eq:kineticMomentum}. This means that a plot of the kinetic momentum spectrum would be shifted by approximately \(2.4 n\ \mathrm{GeV}\) compared to Fig.~\ref{fig:momentumDistributionLHC}. As the width of the peak shown is much greater than this, the approximation of an isotropic kinetic momentum distribution is also justified.

A similar analysis with, for example, the Sauter pulse
\begin{equation}
    B_\mu^{\mathrm{ext}} = B \, \mathrm{sech}^2 \left(\frac{\pi}{2} \omega t\right) \delta_{\mu 2},
\end{equation}
shows a similar momentum distribution, differing only by an \(O(1)\) factor. This suggests that the general structure of the momentum distribution may be a consequence of the localisation of the magnetic field to a time interval of order $1/\omega$, and not of its specific functional form.

\section{Limitations of the free particle approximation} \label{sec:limitations}

The worldline method outlined in Section~\ref{sec:fpa} is based upon two assumptions that are ultimately not satisfied for monopole production in highly relativistic heavy-ion collisions.
These were discussed in Ref.~\cite{Gould:2019myj} where it was judged that the worldline method does not apply at the high energies relevant to monopole searches in modern particle colliders.
However, we will now argue that there is strong evidence that Eq.~\eqref{eq:analyticMomentumDistribution} nevertheless provides a lower bound on the absolute probability of producing monopoles with a given mass. In Section~\ref{sec:uncertainty}, we will also present general arguments for why the relative momentum distribution we have calculated is likely to be accurate to within an \(\mathcal{O}(1)\) factor.

The first assumption made in deriving Eq.~\eqref{eq:analyticMomentumDistribution} is the free particle approximation, in which  worldline self-interactions are neglected. As it is known that if monopoles exist they must carry magnetic charge $g$ in units of \(g_D \approx 20.7 \gg 1\), these self-interactions are expected to significantly modify the result. However, we argue that these strong coupling effects should only enhance the monopole production probability at all momenta, implying that experimental searches in heavy-ion collisions can test the existence of magnetic monopoles in a computable mass range.

In Ref.~\cite{Gould:2019myj} we moved beyond the FPA and included (resummed) leading order corrections due to worldline self-interactions for the inclusive cross section. This removes the requirement that \(g\) must be small, and replaces it with the condition
\begin{equation}
    n \gamma \ll \frac{8 Z e^2}{\pi^2} \approx 6
\end{equation}
for lead ions. This is also not satisfied for the highly relativistic collisions that occur in modern particle colliders, but the results of Ref.~\cite{Gould:2019myj} suggest that corrections due to the strong coupling of the monopoles enhance the probability of production. This was shown to be true not just to leading order, but also to all orders for parameter values such that \(m \omega / g B \sim 1\). Unfortunately, as previously discussed, for present-day heavy-ion collisions we require \(m \omega  / g B \gg 1\), a region of parameter space we were unable to access numerically, as this introduces a difficult-to-resolve hierarchy of scales.

There is still, however, reason to believe that strong coupling effects enhance monopole pair production even in the strongly time-varying fields in heavy-ion collisions.
This can be seen by examining the shape of the worldlines for \(m \omega / g B \gg 1\): the elliptical worldline solutions computed in Ref.~\cite{Gould:2019myj} become increasingly prolate in the imaginary time direction as \(m \omega / g B\) increases.
For very strongly varying fields, the majority of the worldline instanton consists of a pair of worldlines almost parallel to the imaginary time axis; the monopoles are nearly stationary.
Indeed, Eq.~\eqref{eq:analyticMomentumDistribution} with \(\vec{p} = 0\) may be obtained by considering a monopole-antimonopole pair that are created at \(t = -i/\omega\) and remain stationary until \(t = +i/\omega\), when they are destroyed.
The action of a stationary monopole-antimonopole pair is precisely what we studied in Ref.~\cite{ho2019classical}.
The calculation in that paper was carried out using lattice field theory techniques, valid to all orders in the monopole coupling.
We found that for 't Hooft--Polyakov monopoles, the action of a stationary monopole-antimonopole pair, where the attractive force between them is balanced by an external field, is lower than if the pair were considered to be point particles interacting by Coulomb forces.
This holds until the magnetic field becomes so strong that Schwinger production is expected to become unsuppressed; in Ref.~\cite{ho2019classical} we found that 't Hooft--Polyakov monopoles would be produced by a classical instability if external field strengths reach this magnitude.

In addition, in Ref.~\cite{Ho:2021uem} we computed the field-theoretic instanton for Schwinger production of solitonic monopoles, taking the full spacetime dependence of the instanton into account but neglecting the spacetime dependence of the external field. We again found a universal enhancement of the monopole pair production rate compared to the FPA up to the critical field strength.

The results of Refs.~\cite{ho2019classical, Ho:2021uem} also help to address the second limitation of the worldline method: the fact that it assumes monopoles are point particles. This assumption is only valid if the minimal radius of curvature of the worldline is large compared to the classical monopole size, an approximation that was shown to break down in Ref.~\cite{Gould:2019myj} for ultrarelativistic heavy-ion collisions.
The calculation of Refs.~\cite{ho2019classical, Ho:2021uem} was performed with composite monopoles, resolving their internal structure, and so moves beyond the small-monopole assumption. As previously discussed, the results of this calculation suggest that the true production probability is enhanced compared to Eq.~\eqref{eq:analyticMomentumDistribution}.

The above arguments show that, at least for 't Hooft--Polyakov monopoles, the effects of finite size and strong coupling are only expected to enhance monopole production.
For production of such monopoles, Equation~\eqref{eq:analyticMomentumDistribution} is expected to give a lower bound on the production probability.
Combining this with the known classical trajectory of the produced monopoles, it is possible to place a lower bound on the number of monopoles of a given mass that will be detected following a run of heavy-ion collisions, if indeed such monopoles exist.
Either such monopoles are observed, or their nonobservation in turn can be used to give a lower bound on the monopole mass.

One complication that is currently not resolved by any known calculation of monopole production in heavy-ion collisions is the issue of whether the monopole ``fits'' in the magnetic field.
While the magnetic field is extended over a region of order $R$ in the $x$ and $y$-directions, in the \(z\) direction it is much narrower due to Lorentz contraction along the beam axis with a width of order \(R / \gamma\). Taking the classical radius of the monopole \(r_\mathrm{cl} = g^2 / 4 \pi m\), this suggests that monopoles with \(m < \frac{\gamma g^2}{4 \pi R}\) are large compared to the extent of the magnetic field in the $z$-direction. 
In the 5.02 TeV lead-lead collisions at the LHC, this corresponds to a mass of $2600~{\rm GeV}.$
It is not yet known how to account for this effect.

The limited spatial extent of the field in the \(z\) direction also has consequences for our computation of the distribution of the \(p_z\) component of the monopole momentum. Our calculation in Sec.~\ref{sec:fpa} was performed in the background of the spatially homogeneous field~\eqref{eq:heavyIonMagneticField}, which means that the dual potential may be written as a function of time only, and the real part of the \(z\) coordinate along the worldline remains constant. In general the worldline instanton solutions in the background of the full fields~\eqref{eq:fit_functions} are not yet known, though due to the $z\to -z$ symmetry the \(p_z = 0\) family of solutions must be identical to the corresponding solutions we have found. For \(p_z \neq 0\), however, we expect the instanton solutions in the background of the full field to deviate along the real \(z\) axis. This suggests that the relative probability distribution of such solutions will differ from our result~\eqref{eq:final_result}, though the difference can be at most of order $\mc{O}(1)$ as no new scales enter the problem.

\section{Monopole momenta from the uncertainty principle} \label{sec:uncertainty}

The relative momentum distribution of Eq.~\eqref{eq:final_result} was obtained using several approximations, most importantly the neglect of monopole-antimonopole interactions (the FPA) and finite size effects.
The arguments of Sec.~\ref{sec:limitations} suggest that including such corrections is likely to increase the overall monopole production probability.
However, those arguments have little to say about the relative momentum distribution.

Interestingly one can see that the width of the distribution saturates the time-energy uncertainty principle~\cite{aHeisenberg:1927zz,Bauer:1978wd}, which suggests that the results may be generally valid, at least as a lower bound on the momentum variance.

The time-energy uncertainty principle does not have the same rigorous basis in quantum mechanics as the uncertainty principles for conjugate observables, because time is not represented by an operator. Therefore the general statement
\begin{equation}
\Delta t \Delta E\ge   \frac{\hbar}{2}  
\label{equ:uncertainty}
\end{equation}
has many different interpretations and explicit formulations. It is most commonly discussed as representing uncertainty of measurement outcomes, but the aspect that is the most relevant for our current discussion is the {\it preparation time-energy uncertainty}, which is also on a more solid theoretical footing than many other formulations~\cite{Busch2008}.

The preparation time-energy uncertainty refers to any process of preparing a quantum system to a state with a definite energy $E$, where the system is assumed to have a continuous spectrum. The statement is that if the preparation process is carried out in a finite time $\Delta t$, then it is impossible to achieve the desired energy with absolute precision. Instead the energy of the final state of the quantum system will have uncertainty of at least $\Delta E$ given by Eq.~(\ref{equ:uncertainty}).

This corresponds to the situation in our setup. The quantum fields are initially in their vacuum state, and are then exposed to an external perturbation in the form of the time-dependent background magnetic fields. This perturbation, which lasts for a time $\Delta t\approx 1/\omega$, can be considered as the preparation process. With some finite probability, it takes the system to a final state consisting of a monopole-antimonopole pair, with energy
\begin{equation}
    E=2\sqrt{m^2+|\mathbf{p}|^2}=2(m+E_{\rm kin}),
\end{equation}
where $E_{\rm kin}=\sqrt{m^2+|\mathbf{p}|^2}-m$ is the kinetic energy of a single monopole.
This statement does not depend on any assumptions about the nature of the monopoles or approximations. 

In order for the monopoles to be produced at rest, this energy would have to be exactly $E=2m$.
However, the uncertainty principle then tells us that it has uncertainty $\Delta E\ge \omega/2$, and therefore the typical kinetic energy must satisfy the relation
\begin{equation} \label{eq:kinetic_energy_constraint}
    \langle E_{\rm kin}\rangle \gtrsim \frac{\omega}{4}.
\end{equation}
Comparing this with the predicted relative momentum distribution (\ref{eq:final_result}), which we can write as
\begin{equation}
    n_{\rm rel}=\exp\left(-\frac{4E_{\rm kin}}{\omega}\right),
\end{equation}
we can see that it saturates this bound. 
This suggests that it should be a good approximation even when the specific assumptions made in the calculation are not valid:
Even though the precise shape of the exact relative momentum distribution can be different, its width cannot be any narrower without violating the uncertainty principle.
And conversely, if the finite preparation time is the dominant factor in determining the width of the distribution, then we do not expect it to be any wider either.

This observation may also explain why the momentum distribution (\ref{eq:final_result}) is approximately isotropic, even though the external field of Eq.~\eqref{eq:heavyIonMagneticField} defines a preferred direction in space. It is because it is primarily determined by the quantum uncertainty due to the short duration of the production process, rather than any details of that process such as the direction of the field.

The position-momentum uncertainty principles impose
constraints on the momentum distribution, as well.
In the $x$ and $y$ directions, the electromagnetic field varies over a distance $\mc{O}(R)$, the ion radius,
hence the corresponding uncertainty relations imply that $\Delta p_x, \Delta p_y \gtrsim 1/R$.
For ultrarelativistic collisions, this is a much weaker condition than Eq.~\eqref{eq:kinetic_energy_constraint}, so it does not alter the momentum distribution.
In the $z$-direction, the electromagnetic field extends over a distance $\Delta z \approx 1/\omega$, leading to a constraint on $\Delta p_z$ of the same order as Eq.~\eqref{eq:kinetic_energy_constraint}.
While this constraint on $p_z$ is satisfied by our result \eqref{eq:final_result} at the level of orders of magnitude, a more complete calculation including the $z$-dependence may lead to an $\mc{O}(1)$ correction to the shape of the momentum distribution in the $z$-direction, and hence to an $\mc{O}(1)$ anisotropy in this direction.

Finally, we note that uncertainty principles can only constrain the relative momentum distribution, not the absolute monopole-antimonopole pair production probability, for which the two approximations, LCFA and FPA, predict quite different values (see Appendix~\ref{appendix:total_cross_sections}).
As argued in Sec.~\ref{sec:limitations}, there are strong indications that both of these approximations underestimate the absolute production probability.
However, significant theoretical uncertainties remain, and the discrepancies between these approximations give some guide as to their magnitude.
For a conservative estimate of the full probability distribution, it would therefore be sensible to use the lower of the two predictions for the overall normalisation, but the relative momentum distribution could be obtained using the FPA worldline calculation.
For a discussion of the momentum distribution in the LCFA, see Appendix~\ref{appendix:lcfa}.

\section{Conclusions} \label{sec:conclusions}

In this work, we have estimated the momentum distribution of magnetic monopoles produced by the dual Schwinger effect in the strong magnetic fields of ultrarelativistic heavy-ion collisions.
Our main result, Eq.~\eqref{eq:final_result}, gives an explicit formula for this momentum distribution calculated within the FPA.
While this approximation misses contributions due to monopole-antimonopole interactions, the general arguments of Secs.~\ref{sec:limitations} and \ref{sec:uncertainty} imply that such contributions cannot qualitatively modify the result, except for a possible $\mc{O}(1)$ inhomogeniety in the $z$-direction.
In particular a more narrow momentum distribution than what we have found would be in conflict with the preparation time-energy uncertainty principle.

These results can be applied directly to future searches for magnetic monopoles in heavy-ion collisions, such as the MoEDAL experiment at the LHC, for which the potentially large Standard Model backgrounds in heavy-ion collisions do not pose a problem.
More generally, we would like to reiterate the importance of performing monopole searches in heavy-ion collisions.
This is because the strong, classical magnetic fields in heavy-ion collisions are able to overcome the usual exponential form factor suppression of the production cross section~\cite{Witten:1979kh,Drukier:1981fq}.
Thus, searches in heavy-ion collisions (unlike in proton or $e^-e^+$ collisions) have the possibility to conclusively test, in a model-independent way, the existence of magnetic monopoles in a given mass range.

Regarding the mass reach of monopole searches in heavy-ion collisions, some preliminary estimates can be made using the total cross sections of the LCFA and the FPA.
In the LCFA, the total cross section is very strongly mass dependent, and the mass reach of LHC heavy-ion collisions is approximately $\sqrt{g^3B/(4\pi)} \sim 70n^{3/2}$~GeV.
In the FPA, depending on the precise experimental acceptance, the mass reach is $\mc{O}(100~\mr{GeV})$.
These masses are at least an order of magnitude lighter than typical expectations for solitonic monopoles, but elementary monopoles may in principle have any mass.
While significant theoretical work remains to calculate the true production cross section, both these approximations are expected to provide lower bounds on the true cross section~\cite{Gould:2019myj}.
Thus monopole searches in heavy-ion collisions at the LHC have the potential to conclusively test the existence of magnetic monopoles with masses of $\mc{O}(100~\mr{GeV})$ for the first time~\cite{Gould:2017zwi}.

\section*{Acknowledgements}
The authors would like to thank Cesim Dumlu, Gerald Dunne, Christian Kohlf\"{u}rst, Igor Ostrovskiy and Aditya Upreti for useful discussions.
The authors were supported by the Research Funds of the University of Helsinki, and U.K.~Science and Technology Facilities Council (STFC) Consolidated grants ST/P000762/1, ST/T000732/1 and
 ST/T000791/1. AR was also supported by an Institute for Particle Physics Phenomenology Associateship.

\appendix

\section{The locally constant field approximation} \label{appendix:lcfa}

The locally constant field approximation (LCFA) is the leading order in an expansion in derivatives of the electromagnetic field.
Within this approximation it is possible to include the monopole-antimonopole interactions to all orders in the coupling~\cite{Affleck:1981ag,Affleck:1981bma,Ho:2021uem}, and hence the LCFA is complementary to the FPA which we have used in Sec.~\ref{sec:fpa}.
However, the LCFA is not expected to be reliable for calculating the momentum distribution of monopoles produced in ultrarelativistic heavy-ion collisions.
This is because, as argued in Sec.~\ref{sec:uncertainty}, the momenta of the produced monopoles are chiefly due to the fast varying of the external electromagnetic field.
Nevertheless, we include the calculation of the momentum distribution in the LCFA here, as it sheds some light on the orders of magnitude of various contributions to the momentum spectrum.

The LCFA has been widely used to study the production of electron-positron pairs in strong electromagnetic fields.
The earliest discussion of this goes back to Keldysh~\cite{keldysh1965ionization}, who recovered the constant field result in the limit that the Keldysh parameter goes to zero $\xi\to 0$.
For more recent discussion of the LCFA in the context of the Schwinger production of electron-positron pairs, see Refs.~\cite{narozhny2004e+,Gavrilov:2016tuq,Aleksandrov:2018zso}.

In the LCFA the rate of pair production at any given spacetime point is given by the expression in a constant electromagnetic field evaluated at that point.
In our case, the field, Eq.~\eqref{eq:fit_functions}, satisfies $\mb{E}\cdot\mb{B}=0$ everywhere, and hence one can always perform a boost (along the $z$-axis) to set $\mb{E}=\mb{0}$ at a given spacetime point.
In the new frame, there is only the magnetic field, and hence the pair production probability is given by Eq.~\eqref{eq:affleck_manton}.
Once produced, the charged pairs are then evolved classically through the electromagnetic field, according to the (dual) Lorentz force law.
Thus the final momentum distribution in the LCFA has three distinct contributions:
\begin{itemize}
    \item[(i)] the boost to the frame with $\mb{E}=\mb{0}$,
    \item[(ii)] the initial momentum spread in quantum tunnelling,
    \item[(iii)] the classical time evolution.
\end{itemize}
Note that in the FPA, points (ii) and (iii) are accounted for together by the use of the canonical momentum; see Eq.~\eqref{eq:kineticMomentum}.
Further, point (i) is neglected in Sec.~\ref{sec:fpa} as the $z$-dependence of the field is dropped in Eq.~\eqref{eq:heavyIonMagneticField}.

The momentum distribution of particles produced by a constant field was found by Nikishov~\cite{Nikishov:1969tt}.
In the directions transverse to the magnetic field, the result is captured simply by a shift of the form $m^2 \to m^2 +\ptrans^2$, where $\ptrans$ denotes the transverse momentum, c.f.\ Eq.~\eqref{eq:analyticMomentumDistribution}.
The distribution of the longitudinal momentum, parallel to the external field, $\plong$, depends on the lifetime of the field.
For a constant field existing for a long time, $T$, the distribution is very close to flat, extending to $\pm \int gB dt = \pm gBT$.
This can be understood as monopoles being produced with zero longitudinal momentum, uniformly in time, and then accelerated by the field.
This observation forms the basis of the treatment in the LCFA, in which particles are considered to be produced with zero longitudinal momentum by quantum tunnelling, in the vicinity of some spacetime point.
This approach to the longitudinal momentum distribution misses transitory phenomena related to the turning on and off of the field~\cite{Aleksandrov:2018zso}, interference effects~\cite{Kohlfurst:2017git} as well as corrections related to the Heisenberg uncertainty principle~\cite{Garriga:2012qp,Garriga:2013pga}, c.f.\ also Sec.~\ref{sec:uncertainty}.
However, for sufficiently weak and slowly varying fields it approximates well the correct momentum distribution.

Combining the contributions (i) and (ii) to the momentum distribution we find the initial momentum distribution, before the classical time evolution,
\begin{align} \label{eq:lcfa_distribution}
n_{\rm}(t,\mb{x},\mb{p}) =& \frac{g \mc{F} }{(2\pi)^3}\exp\left(-\frac{\pi m^2}{g\mc{F} } + \frac{g^2}{4}\right)
\nonumber \\
&\times \exp\left(-\frac{\pi}{g\mc{F} } \left(\ptrans -  \mb{q}_{\perp}\right)^2 \right) \delta \left(\plong\right),
\end{align}
where,
\begin{align}
\mc{F} &= \F ,&
\mb{q}_{\perp} &= \frac{m|\mb{E}|}{\F}\hat{\mb{E}}\times\hat{\mb{B}},
\end{align}
and the hats denote unit normalised vectors.
This formula is valid for sufficiently slowly varying fields, and assuming $\mb{E}\cdot\mb{B}=0$.
For monopoles with spin $s$ there is an overall multiplicative factor of $(2s+1)$, and the sum of the spins of the produced pair must equal zero.

Inserting the specific form of the fields into Eq.~\eqref{eq:lcfa_distribution}, we find an initial probability distribution that is peaked in spacetime around $(z,t)=(0,0)$, with a width of size $\omega^2 \Delta t^2 \sim \omega^2 \Delta z^2 \sim g B/(3 \pi  m^2)$.
The boost at nonzero values of $z$ which is needed to set the electric field to zero induces a corresponding momentum of order $p_z/m \approx 3 \omega^2 z t \sim gB/(\pi m^2)$.
The additional spread in the transverse momenta which is due to the tunneling process is of this same order, $\Delta p_z/m \sim \Delta p_x/m \sim g B/(\pi m^2)$.
The process of pair production is assumed to be semiclassical, and hence the exponent of Eq.~\eqref{eq:lcfa_distribution} is large and negative.
As a consequence, $g B/(\pi m^2) \ll 1$ and the initial momenta are nonrelativistic.

This initial distribution of magnetic monopoles is then evolved in time according to the dual Lorentz force law,
\begin{equation}
\frac{d p^\mu}{d\tau} = g \tilde{F}^{\mu\nu} p_\nu ,
\end{equation}
where $\tilde{F}^{\mu\nu}=\frac{1}{2}\epsilon^{\mu\nu\rho\sigma}F_{\rho\sigma}$ is the dual electromagnetic field tensor and $\tau$ is the proper time along the particle worldline.
The monopole-antimonopole pairs will be accelerated in opposite directions by the magnetic field, and their trajectories will be curved by the electric field.
For monopoles with nonzero spin, there will also be an additional term coupling the resulting dipole moment to the external field; (see for example Ref.~\cite{Kohlfurst:2017git}).
The final momentum distribution is achieved after the monopoles have left the interaction region, at $t=\infty$ on the celestial sphere.

The final momentum distribution predicted within the LCFA is complicated, but we can extract some general features.
To understand the magnitude of contribution (iii) to their final momenta, we consider the simple case of a monopole-antimonopole pair produced at the origin, and then accelerated apart by the magnetic field.
Their asymptotic momenta are given by
\begin{align} \label{eq:lorentz-force-momentum}
    p_y = \pm g \int_0^\infty \mr{d}t\ B_y = \pm \frac{gB}{\omega} \approx \pm \frac{Z e g}{2\pi R}.
\end{align}
Note that while a stronger magnetic field is produced at larger centre-of-mass energies, the field lasts for a correspondingly shorter time, in such a way that the final result is independent of the centre-of-mass energy.
For lead ions Eq.~\eqref{eq:lorentz-force-momentum} gives $p_y\approx \pm 2.4n~\mr{GeV}$.
As the peak of the initial monopole production is at the origin, and at zero initial momentum, the final momentum distribution will have two peaks at $p_y\approx \pm 2.4n~\mr{GeV}$.
If we consider particles starting at nonzero velocities instead of at rest, the same conclusions will hold regarding the change of momentum due to the field, as the momentum evolution equations are linear.

In summary, the LCFA predicts a nonrelativistic initial momentum distribution which receives additive corrections from the time evolution, of order a few GeV.
For the ultrarelativistic heavy-ion collisions performed at the LHC, this momentum distribution is much more strongly peaked around zero than the result from the FPA, Eq.~\eqref{eq:final_result}.
As a consequence, for such ultrarelativistic collisions, the prediction of the LCFA violates the preparation time-energy uncertainty principle.
This is not surprising, as the LCFA treats the production process as adiabatic, whereas it is in fact closer to the opposite limiting case, of an instantaneous process.

\section{Total cross sections} \label{appendix:total_cross_sections}

For ease of reference and consistency of notation, we quote here the total cross sections calculated in the LCFA and FPA approximations~\cite{Gould:2019myj}, after integration over the momentum distribution.
These read
\begin{align}
\frac{1}{2\pi b}\frac{d\sigma_{\rm LCFA}}{db} &\approx \frac{(g B)^{4}}{18 \pi^3 m^4 \omega^2\Omega_x\Omega_y}\exp\left(-\frac{\pi m^2}{g B}+\frac{g^2}{4}\right), \\
\frac{1}{2\pi b}\frac{d\sigma_{\rm FPA}}{db} &\approx \frac{(g B)^{4}}{18 \pi^3 m^4 \omega^2\Omega_x\Omega_y} \exp\left(-\frac{4 m}{\omega}\right),
\end{align}
where $B$ and $\omega$ are functions of the collision energy and impact parameter $b$.
The remaining factors, $\Omega_x\approx 2/b$ and $\Omega_y\approx 1/R$, are the slow inverse decay lengths in the $x$ and $y$ directions.

\bibliography{refs}

\end{document}